\documentclass[a4paper,fleqn]{cas-dc} 

\usepackage[numbers,sort&compress]{natbib}

\usepackage{natbib}

\usepackage[capitalise]{cleveref}
\creflabelformat{equation}{#2\textup{#1}#3}
\Crefname{equation}{Eq.}{Eqs.}
\Crefname{figure}{Fig.}{Figs.}

\begin{document}
\let\WriteBookmarks\relax
\def\floatpagepagefraction{1}
\def\textpagefraction{.001}

\shorttitle{}    

\shortauthors{}  

\title [mode = title]{Liquid xenon positron target flow optimizations}  



%

\author[1]{Max Varverakis}[orcid=0000-0003-4338-924X,]
\cormark[1]
\credit{Software, Formal analysis, Investigation, Data Curation, Visualization, Writing - Original Draft, Writing - Review \& Editing}


\fnmark[1]

\ead{maxvarverakis@physics.ucla.edu}



\affiliation[1]{organization={{SLAC} National Accelerator Laboratory},
            addressline={2575 Sand Hill Road}, 
            city={Menlo Park},
            postcode={94025}, 
            state={CA},
            country={USA}}

\author[1]{Spencer Gessner}[orcid=0000-0002-9713-1116]
\cormark[2]
\credit{Conceptualization, Supervision, Project administration, Writing - Review \& Editing, Funding acquisition}


\ead{sgess@slac.stanford.edu}



\author[2]{Silviu Covrig Dusa}[orcid=0000-0001-9117-8493]
\affiliation[2]{organization={Thomas Jefferson National Accelerator Facility},
            addressline={12000 Jefferson Avenue}, 
            city={Newport News},
            postcode={23606}, 
            state={VA},
            country={USA}}
\credit{Software, Formal analysis, Investigation, Data Curation, Visualization, Writing - Review \& Editing}

\author[2]{Joseph Grames}[orcid=0000-0001-9390-8752]
\credit{Writing - Review \& Editing}

\cortext[1]{Corresponding author}
\cortext[2]{Principal corresponding author}

\author[1]{Nathan Majernik}[orcid=0000-0001-9977-0248]
\credit{Formal Analysis}

\author[3]{Sanjeev Chauhan}
\affiliation[3]{organization={Duke University},
            addressline={2080 Duke University Road}, 
            city={Durham},
            postcode={27708}, 
            state={NC},
            country={USA}}
\credit{Software}

\fntext[1]{Present address: Department of Physics \& Astronomy, University of California, Los Angeles, 90095, CA, USA}


\begin{abstract}
In a previous work, we explored liquid xenon (LXe)-based positron targets as an alternative to conventional high-Z metallic targets for high energy physics applications. In this paper, we apply computational fluid dynamics simulations to the LXe target and containment vessel, including entrance and exit windows. We design a flow geometry that is able to quickly move heated LXe from previous beam pulses out of the target volume without significant boil-off.
While our simulations indicate that the temperature of the exit window can be kept stable, the peak energy deposition density exceeds the widely accepted damage threshold of 35 Jg$^{-1}$ for ILC-like conditions.
\end{abstract}





\maketitle


\section{Introduction}


Conventional solid metal positron targets tend to degrade over time due to the thermal shocks induced by intense electron beams~\cite{Bharadwaj2001}. An ILC-type collider calls for positron production rates on the order of $10^{14}$ positrons per second to reach luminosities of $10^{34}$ cm$^{-2}$s$^{-1}$~\cite{Seimiya2015,Abe2025}. A cost-effective positron source that fulfills this requirement must implement new technologies to mitigate rapid target damage. Hence, the development of positron source technology is a crucial step toward enabling an ILC-type collider~\cite{Chaikovska2022,snowmassPositron,Fukuda:2025}.

Although there have been advancements for solid metal targets, such as water cooling and rotation~\cite{Abe2025,Omori2024}, the added complexity, cost, and upkeep of such a design opens the door for alternative target concepts that are less susceptible to damage.
For example, liquid targets do not degrade in the same manner as traditional solid metal targets. There are multiple candidate substances that have potential applications as future collider particle sources, however many bring their own set of challenges. Liquid lead~\cite{Sheppard2002} and mercury~\cite{Mikhailichenko2006} are not immediately suitable as liquid positron targets due to their toxicity, while GaInSn~\cite{Taylor:2024} is currently being investigated at JLab.

Liquid xenon (LXe) is an attractive target candidate because it is a stable, high-Z, non-toxic liquid at relatively high temperatures. LXe also has a peak energy deposition density (PEDD) threshold of at least 96 Jg$^{-1}$ (based off of the vaporization threshold) which is greater than the 35 Jg$^{-1}$ limit for solid metals~\cite{Ecklund1981,NLC:2001,Burrows:2012,Zang:2014}. A LXe target can therefore handle greater energy deposition than its solid target counterparts and does not degrade over time.

Previous GEANT4~\cite{Geant4} simulations demonstrated that a LXe target can produce positrons with a comparable yield and energy spectrum to conventional metal targets, such as tantalum (Ta) and tungsten-rhenium (W$_{75}$Re$_{25}$)~\cite{Varverakis2023}. Bulk energy deposition estimates indicated that solid beryllium disks were not suitable for LXe positron target windows.

A more detailed LXe positron target design can leverage an optimized flow structure through the target chamber. To achieve such an optimization, we simulate LXe positron production in G4Beamline~\cite{Roberts2007} and obtain energy deposition distributions in the LXe and target windows. We use the energy deposition profiles in tandem with ANSYS Fluent~\cite{ansys_fluent_2025r1,ansys_fluent_2026r2} simulations to develop an optimal flow scheme to limit LXe and window temperatures for an ILC-type beam.

\section{Energy deposition distributions}

The particle simulations presented in this work use a wrapper for the GEANT4 toolkit~\cite{Geant4} called G4Beamline. A 10 cm long ($\sim$3.5 radiation lengths), 7 cm by 7 cm prism of LXe was divided into 245,000 rectangular cells (0.1 cm by 0.1 cm by 0.2 cm) and subjected to $10^3$ electrons at 3 GeV. Energy deposition was obtained in each cell and scaled to the most recent ILC bunch train parameters. After accounting for positron capture losses, we select an operating point of $2.5\times10^{10}$ electrons per bunch as in the parent work~\cite{Varverakis2023}. Both beryllium (Be) and aluminum (Al) disks were tested as windows at thicknesses of 0.5 mm and 2.0 mm. The windows were transversely divided into 0.1 cm by 0.1 cm cells.

\begin{figure}[pos=htbp]
    \centering
    \includegraphics[width=\linewidth]{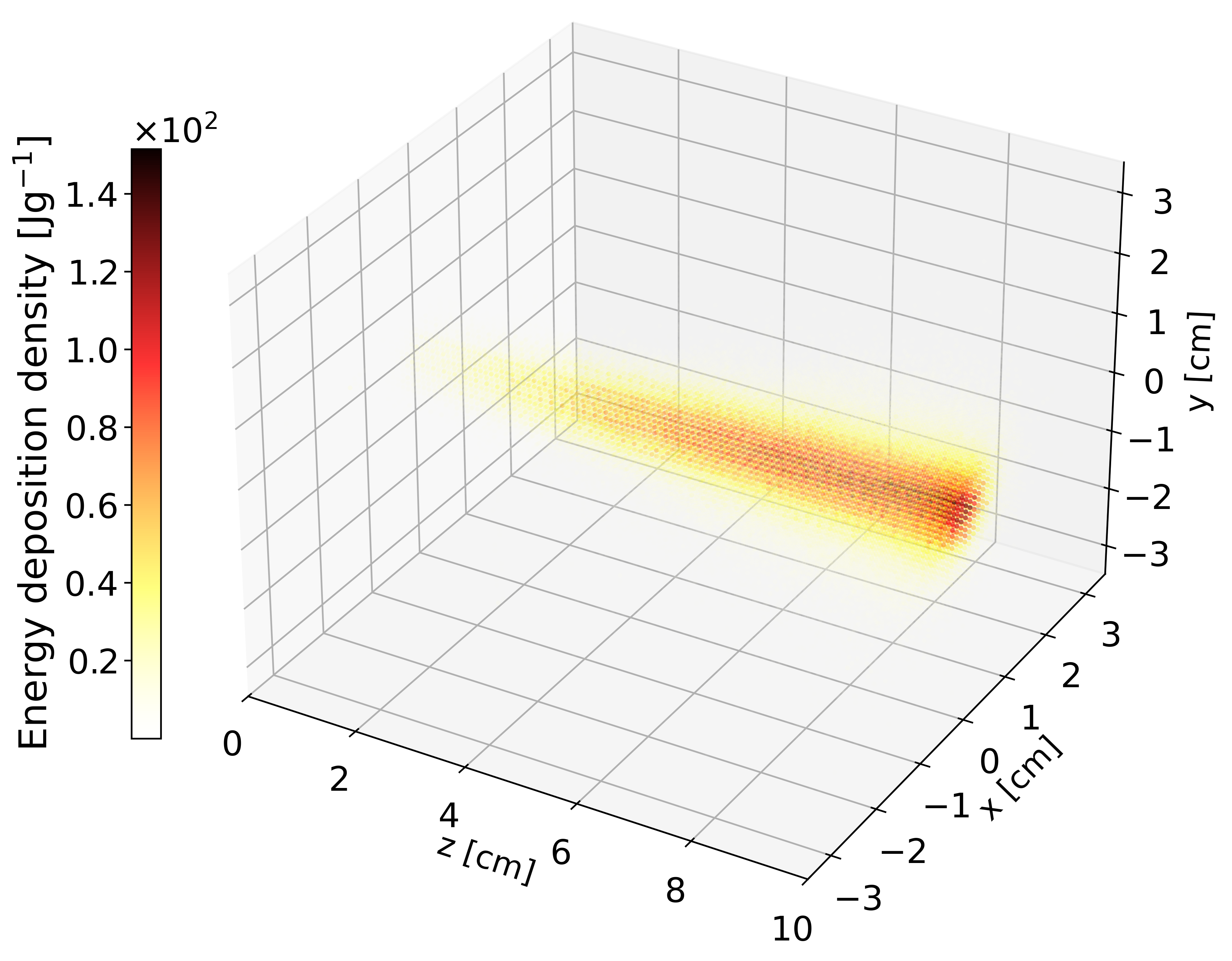}
    \caption{Energy deposition density distribution in the LXe. The beam enters the LXe from the left at $(x,y,z) = (0,0,0)$ after traversing a 0.5 mm-thick Be window. Less than half of one percent of the EDD per LXe cell exceeds the PEDD of 96 Jg$^{-1}$.}
    \label{fig:LXe_scatter_EDD}
\end{figure}

A three-dimensional energy deposition distribution was obtained in the LXe as well as two-dimensional distributions in the entrance and exit windows. \cref{fig:LXe_scatter_EDD} illustrates the energy deposition density ({EDD}) distribution in the LXe per ILC bunch train. Less than half a percent of the energy deposited in the LXe exceeded the PEDD of 96 Jg$^{-1}$, which corresponds to minimal xenon vaporization.

\begin{figure}[pos=htbp]
    \centering
    \includegraphics[width=\linewidth]{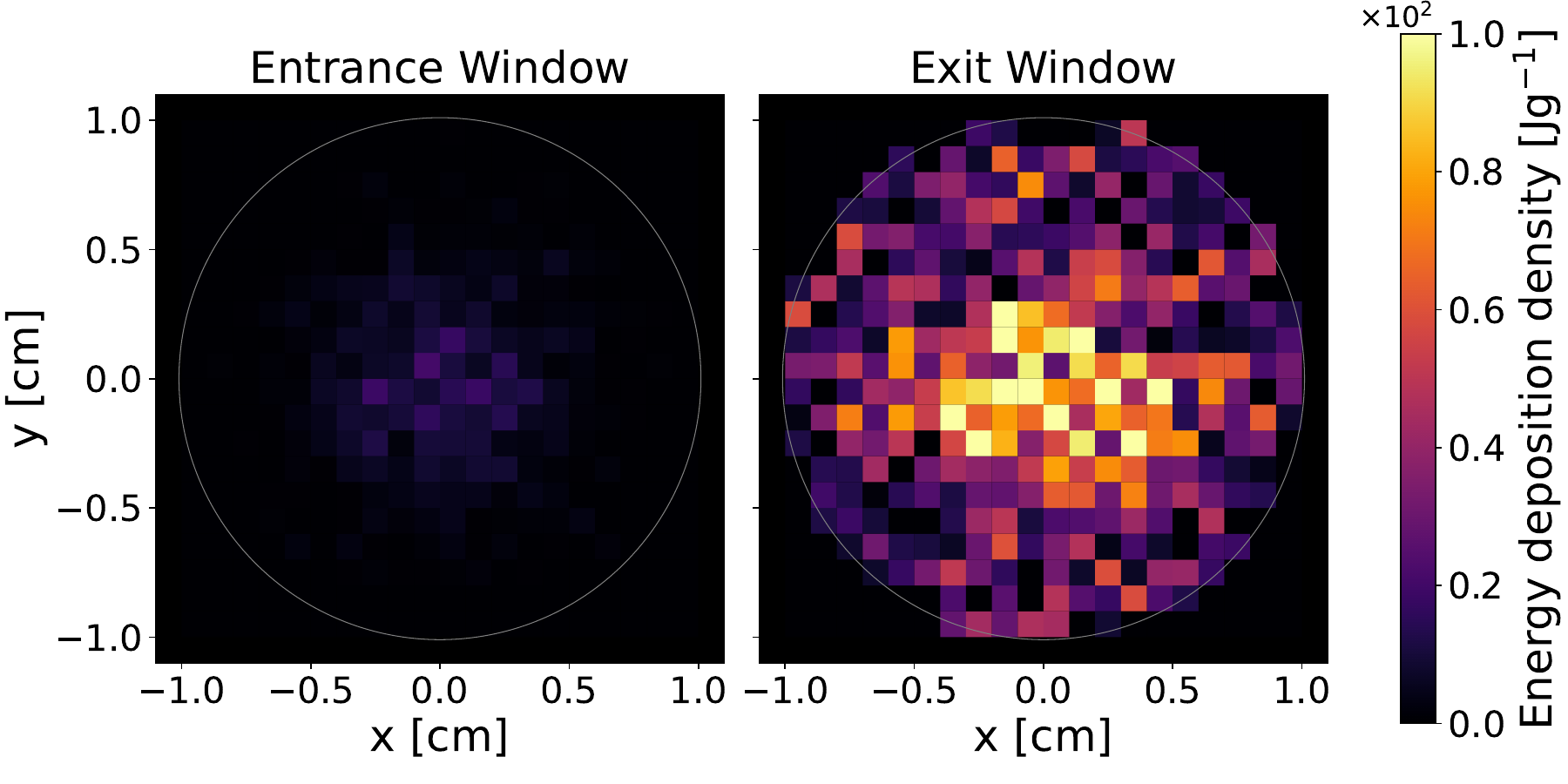}
    \caption{Energy deposition density distribution of 0.5 mm Be windows. The entrance window receives minimal energy deposition whereas the exit window EDD repeatedly exceeds the 35 Jg$^{-1}$ threshold.}
    \label{fig:Be_EDD}
\end{figure}

Both Be and Al exit windows experience significant energy deposition beyond their PEDD (35 Jg$^{-1}$) at both thicknesses. Assuming that this selected PEDD is appropriate for our window materials, this indicates a high risk of material degradation and structural failure due to thermal shocks. The EDD distribution for 0.5 mm-thick Be windows is shown in \cref{fig:Be_EDD}. All four window cases produced similar outcomes. In agreement with~\cite{Varverakis2023}, the entrance windows receive negligible energy deposition.

\section{CFD optimization and results}

The findings from G4Beamline are limited in that they assume a static LXe for the entire ILC bunch train. A key feature of the LXe positron target is that heated xenon can evacuate the IP before the next set of bunches arrive. We use ANSYS Fluent~\cite{ansys_fluent_2025r1,ansys_fluent_2026r2}, a computational fluid dynamics (CFD) software, to study the temporal response of the LXe and windows to an ILC-type beam structure.

\begin{figure}[pos=htbp]
    \centering
    \includegraphics[width=\linewidth]{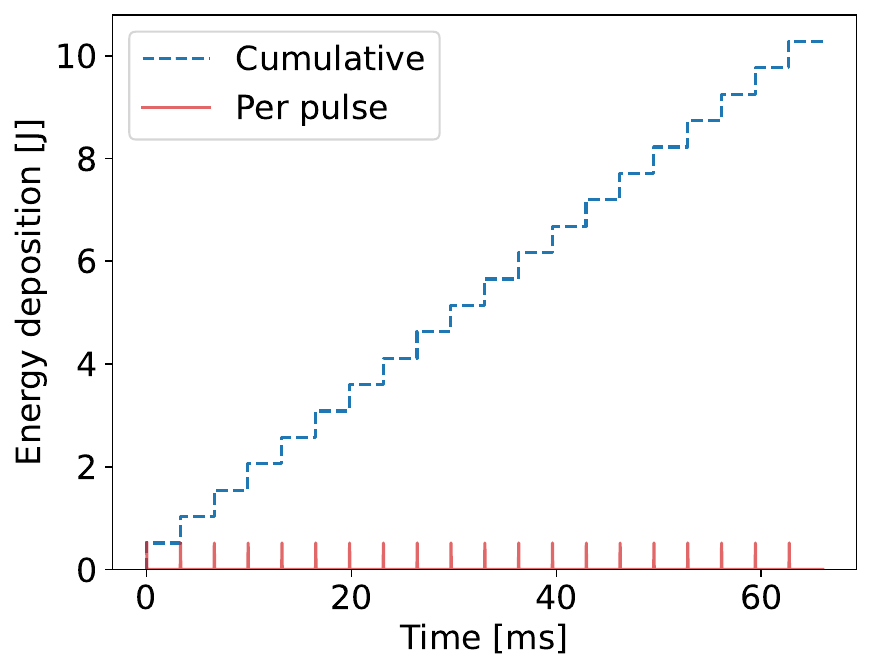}
    \caption{Cumulative (blue dashed line) and per-pulse (solid red line) energy deposition for a 0.5 mm-thick Be exit window subject to an {ILC}-type bunch train derived from G4Beamline simulations. The total energy deposition per beam pulse (66 bunches) is deposited in a 50 $\mu$s time step, followed by 65 50 $\mu$s-steps of no energy deposition. The 3.3 ms interval is repeated 20 times, and then 134 ms elapses with no beam delivery. This time structure is designed to mimic a single {ILC} bunch train delivered to the target. We assume LXe is static over 20 to 50 $\mu$s which results in a similar target response to the actual {ILC} pulse delivery in 474 ns.}
    \label{fig:CFD_Edep}
\end{figure}

Energy deposition data from G4Beamline is used as an input to the CFD simulations and scaled to match the energy deposited from one beam pulse (66 bunches). The bunch train time structure used in our simulations follows the most recent found in the literature~\cite{Omori2024}. Exit window energy deposition resulting from an ILC bunch train is schematically depicted in \cref{fig:CFD_Edep}.

\begin{figure}[pos=htbp]
    \centering
    \includegraphics[width=\linewidth]{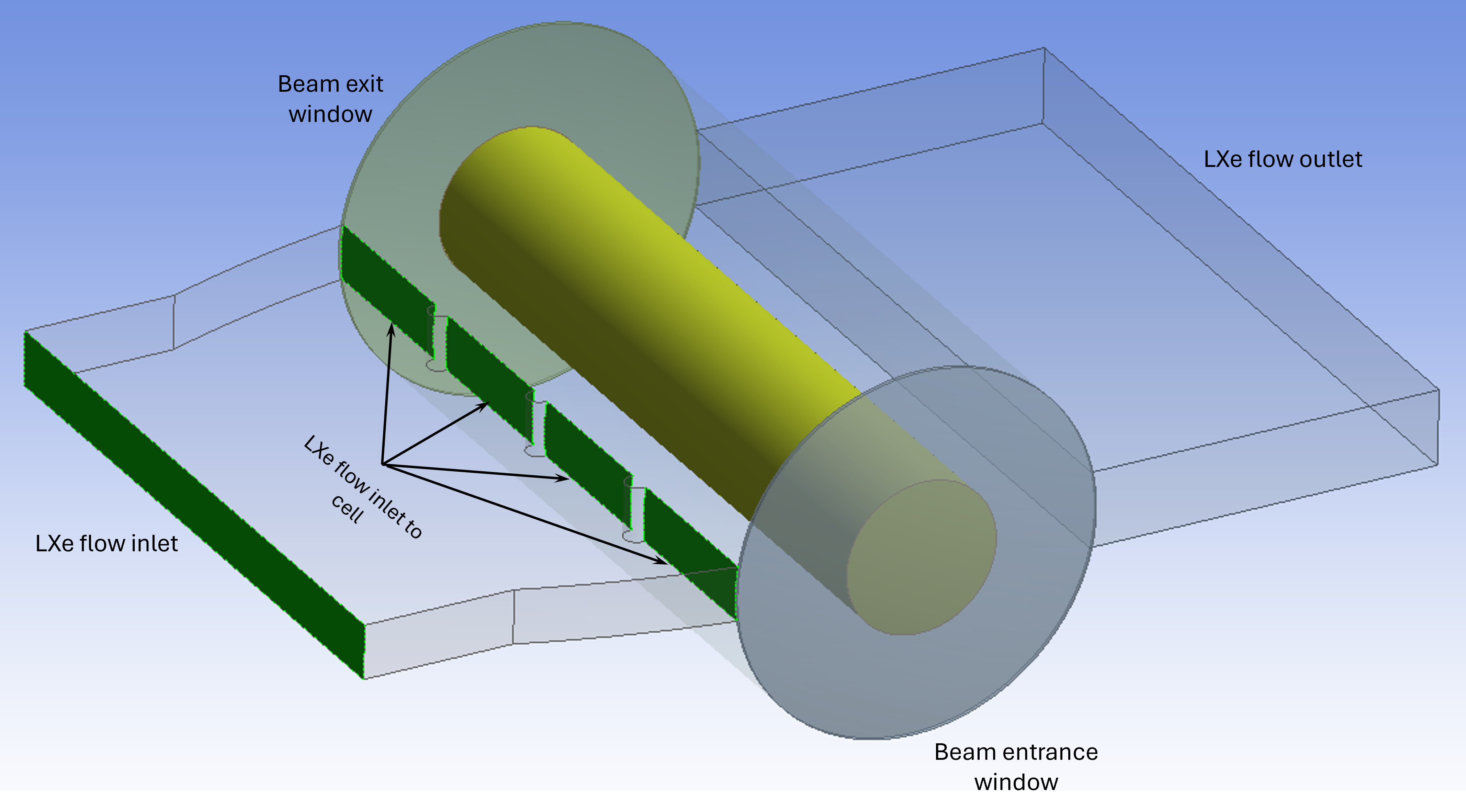}
    \caption{Fluid space model of cell 10. The cell is cylindrical with 4.8~cm diameter and 10~cm length along the beam axis. The cell is designed for transverse flow to the beam axis. The flow inlet areas are highlighted in green. The highlighted yellow volume in the picture is the 1~cm radius cylindrical volume in the cell, centered on the beam axis, which we call the cell's core.}
    \label{fig:Cell10model}
\end{figure}

The CFD driven design focused on (1) optimizing the LXe flow across the beam line in a cylindrical volume inside the target cell with radius 1~cm (called the cell's core), centered on the beam axis, where most of the energy is deposited and (2) mitigating the beam heating of the cell's windows. A total of 10 different cell geometries have been explored in ANSYS Fluent, 8 geometries focused on cylindrical cell shapes, while 2 geometries explored conical cells. The cell geometries were picked to accommodate a full azimuthal symmetry about the beam axis and the ILC requirements for positron capture downstream of the target cell. The last design iteration of the target cell geometry, which we call cell 10, is shown in \cref{fig:Cell10model}. The total volume of the cell for LXe, from the cell flow inlet to outlet is 251~cm$^3$. The target cell is cylindrical in shape with a diameter of 4.8~cm and a length along the beam axis of 10~cm. The most important part of the cell is the flow inlet manifold, which should shape the flow in time and space to mitigate heating in the beam illuminated volume in the cell and optimize convective heat transfer at the cell's windows.

\begin{figure}[pos=htbp]
    \centering
    \includegraphics[width=\linewidth]{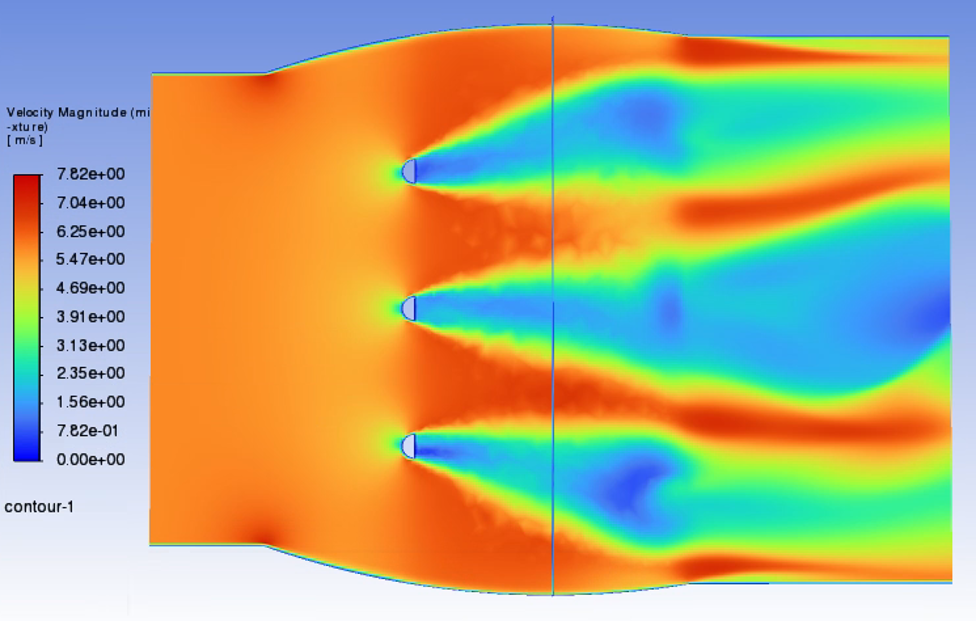}
    \caption{LXe flow structure in cell 10, from ANSYS Fluent simulations. In this top-down view the thin vertical line in the picture is the location of the beam axis. We tailor the LXe flow to optimize the convective heat transfer at the entrance and exit windows.}
    \label{fig:CFD_flow}
\end{figure}

We used a mixture two-phase model with the phase transition (boiling) accounted for in ANSYS Fluent to determine a cell's performance. Material properties have been taken from a NIST database~\cite{NISTdb} and corrected for temperature dependence in isobaric conditions. We considered LXe at 35~psia. At this pressure Xe has a temperature range for the liquid phase between 161.4~K (below which it freezes) and 181.81~K (the saturation temperature). The LXe flows into the cell at 170 K, 35~psia and 10~kg/s. At this mass rate the LXe flow velocity averaged over the inlet area is 5.7 m/s. The LXe flow velocity averaged over the 1~cm cylindrical beam volume in the cell is 3-4~m/s over time. The LXe flow at the cell's windows in the beam interaction region averages over time 5-7~m/s. The LXe pressure loss between flow inlet and outlet of the cell is less than 2.5~psi, which is not a problem for a cryogenic recirculating centrifugal pump. Jefferson Lab has experience operating cryogenic recirculating pumps up to 4~psi of pressure loss. The steady state flow structure is shown in \cref{fig:CFD_flow} in a horizontal cross section through the middle of the cell. In this picture the LXe flows from left to right. Before entering the cell volume the LXe flow is split 4-ways with 3 cylindrical wall inserts, each 3.2~mm diameter. The inserts were designed to optimize the LXe flow in the core of the cell and improve convective heat exchange at the cell's beam line windows.

\begin{figure}[pos=htbp]
    \centering
    \includegraphics[width=\linewidth]{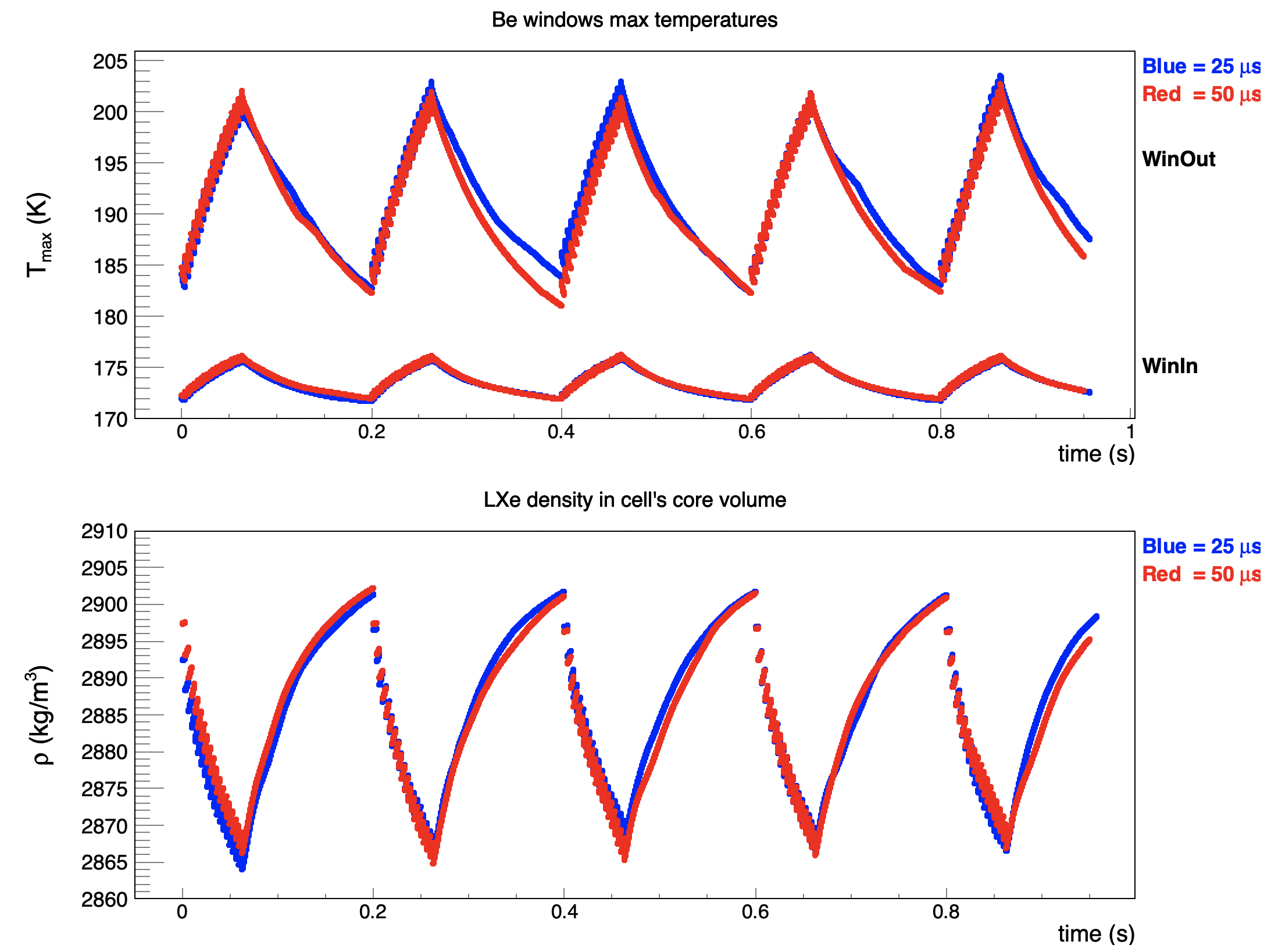}
    \caption{Comparison between using a CFD time step of 25~$\mu$s (blue) and 50~$\mu$s (red). Top plot shows the time evolution of the maximum temperature in the Be windows of the cell, WinOut is the beam exit window of the cell, WinIn is the beam entrance window of the cell. Bottom plot shows the time evolution of the LXe density averaged over the core of the cell, defined as a 1~cm radius cylindrical volume of the cell, centered on the beam axis. The LXe density variation, in the cell's core, between the end of a 20 beam pulse train and the starting of the next 20 beam pulse train is 1.5~\%.}
    \label{fig:CFD_comp25-50}
\end{figure}

\begin{figure}[pos=htbp]
    \centering
    \includegraphics[width=\linewidth]{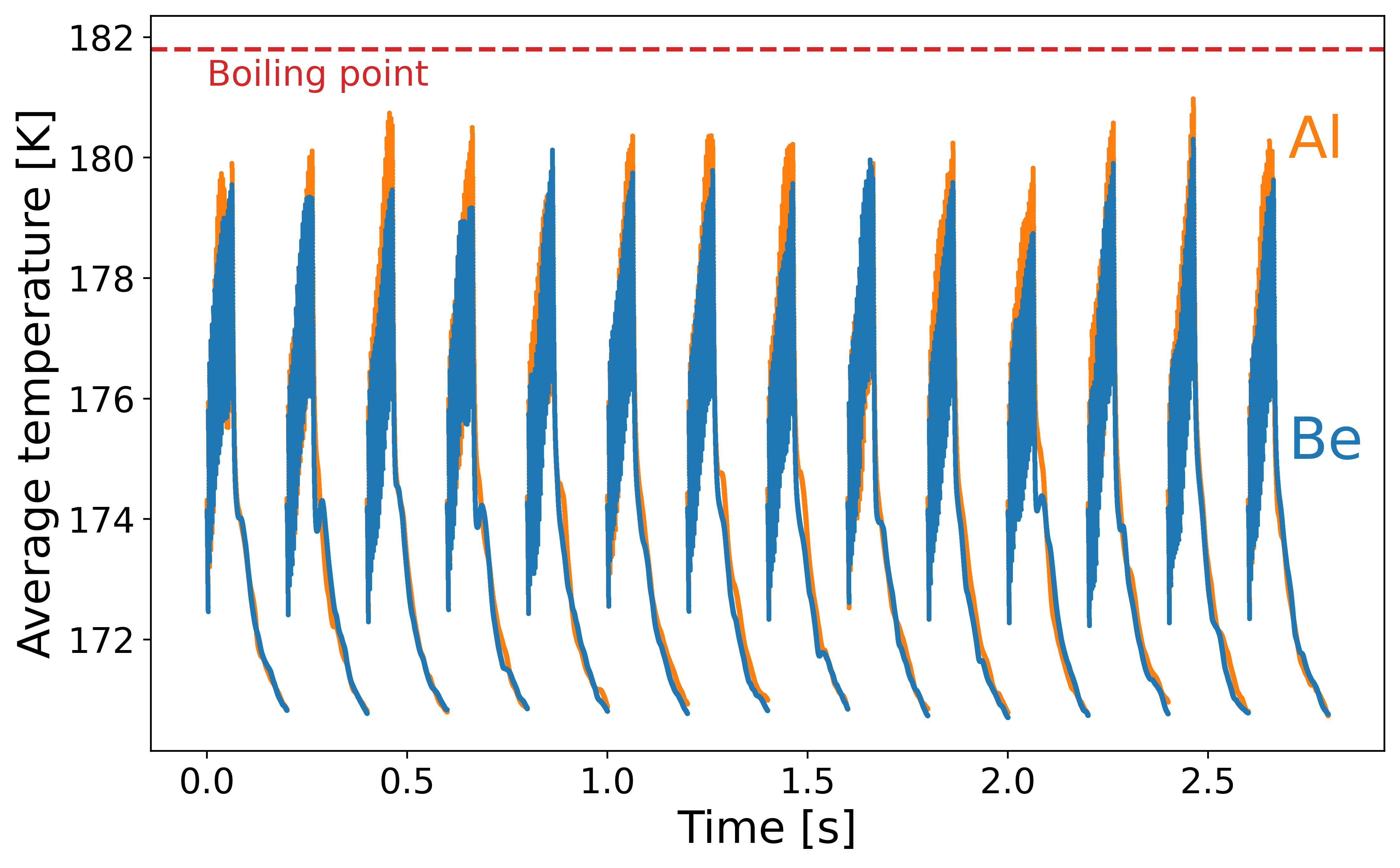}
    \caption{Average temperature in a 1 cm-radius cylindrical column of LXe oriented along the beam axis. Each spike corresponds to one bunch train. Since the energy deposition is concentrated along the beam axis (see \cref{fig:LXe_scatter_EDD}), we see that, on average, the LXe core remains liquid. The LXe boiling point is indicated by the horizontal (red dashed) line.}
    \label{fig:LXe_temp}
\end{figure}

In all the CFD simulations, initially the steady state flow of LXe is established through the cell's geometry, then the beam power deposition, taken directly from G4Beamline, is interpolated to the ANSYS Fluent mesh, separately for the 3 interaction regions: the beam entrance window, LXe and the beam exit window respectively. After the primary beam energy deposition is interpolated, a transient CFD simulation is performed, with a fixed time step, accounting for the ILC beam time structure. The ILC 66 beam bunches last less than 0.5~$\mu$s of the 3.3~ms of a beam pulse. The average mesh size in Fluent is 1~mm, and for a fluid flow in the range of 5~m/s, the fluid traverses a mesh cell in about 0.2~ms, so the fluid is essentially static on the time scale of the ILC beam bunches, of 0.5~$\mu$s and for the average Fluent mesh cell size. We have investigated two time steps for the CFD simulations: 25~$\mu$s and 50~$\mu$s. For the target cell with Be windows there is no significant change in the LXe density and cell windows maximum temperature variations over time between the two time scales, see~\cref{fig:CFD_comp25-50}. Both time steps predict the same performance for the cell. The difference is that using a 25~$\mu$s time steps for CFD simulations takes twice as long to accumulate data for the same period of beam on target compared to using a 50~$\mu$s time step. With a time step of 50~$\mu$s an ILC beam pulse of 3.3~ms is sampled with 66 CFD time steps, enough to predict the effects of the beam on the fluid's behavior. All the CFD results presented here, except for those in~\cref{fig:CFD_comp25-50}, have been obtained by using a 50~$\mu$s fixed time step.

\begin{figure}[pos=htbp]
    \centering
    \includegraphics[width=\linewidth]{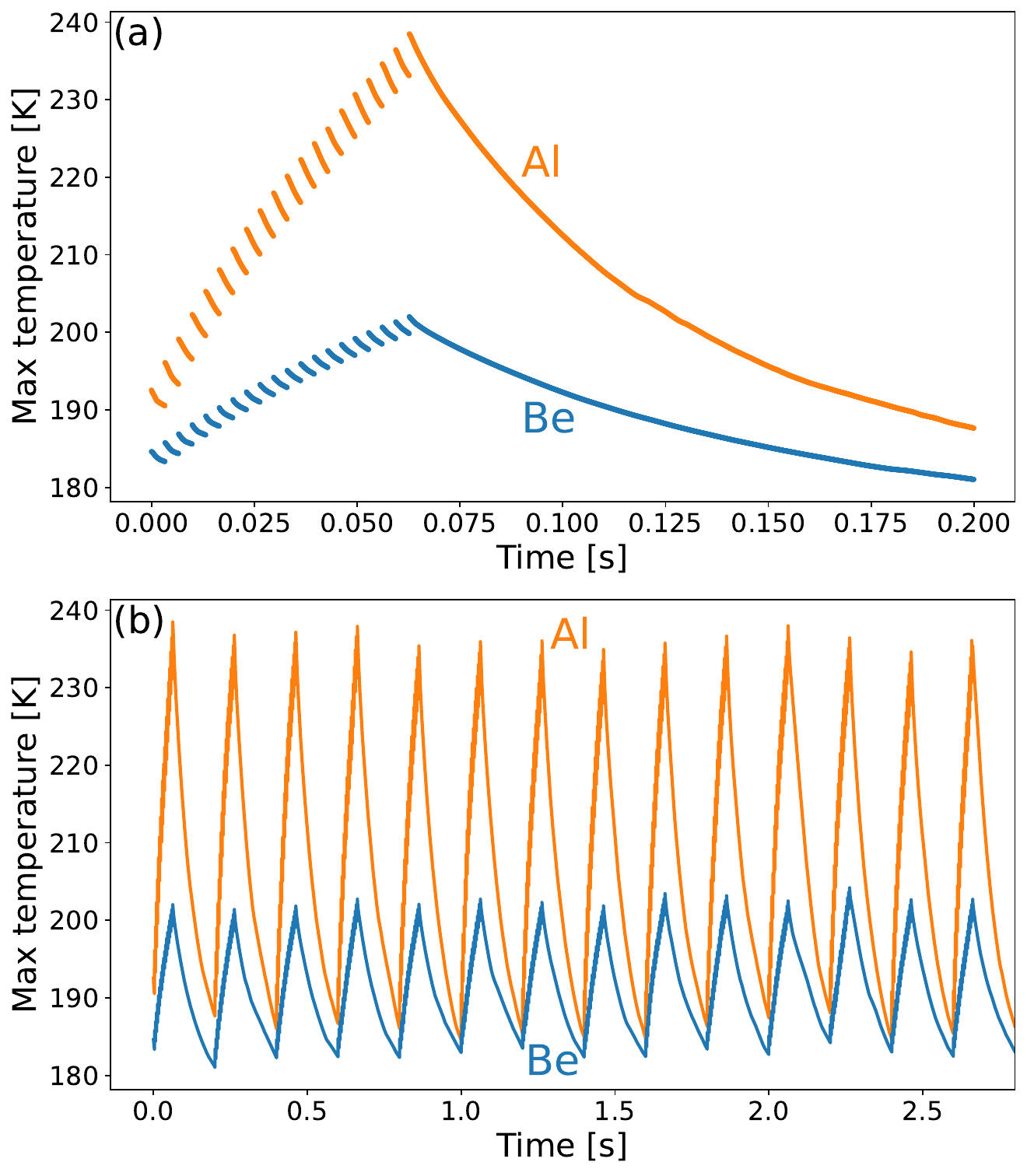}
    \caption{Maximum temperature in 0.5 mm-thick beryllium and aluminum exit windows for (a) one bunch train and (b) multiple bunch trains. The aluminum exit window heats significantly more than beryllium. Each spike in plot (b) corresponds to one bunch train (20 pulses).}
    \label{fig:exit_max_temp}
\end{figure}

Average temperature of the core of LXe over the duration of multiple bunch trains is shown in \cref{fig:LXe_temp}. Thermal steady state is maintained, as evidenced by the temperature nearly returning to baseline before the next bunch train is delivered at the IP. Likewise, the maximum temperature of both beryllium and aluminum exit windows is depicted in \cref{fig:exit_max_temp}. It is clear from the figure that aluminum experiences considerably greater heating than beryllium, likely due to the longer radiation length.

\begin{figure}[pos=htbp]
    \centering
    \includegraphics[width=\linewidth]{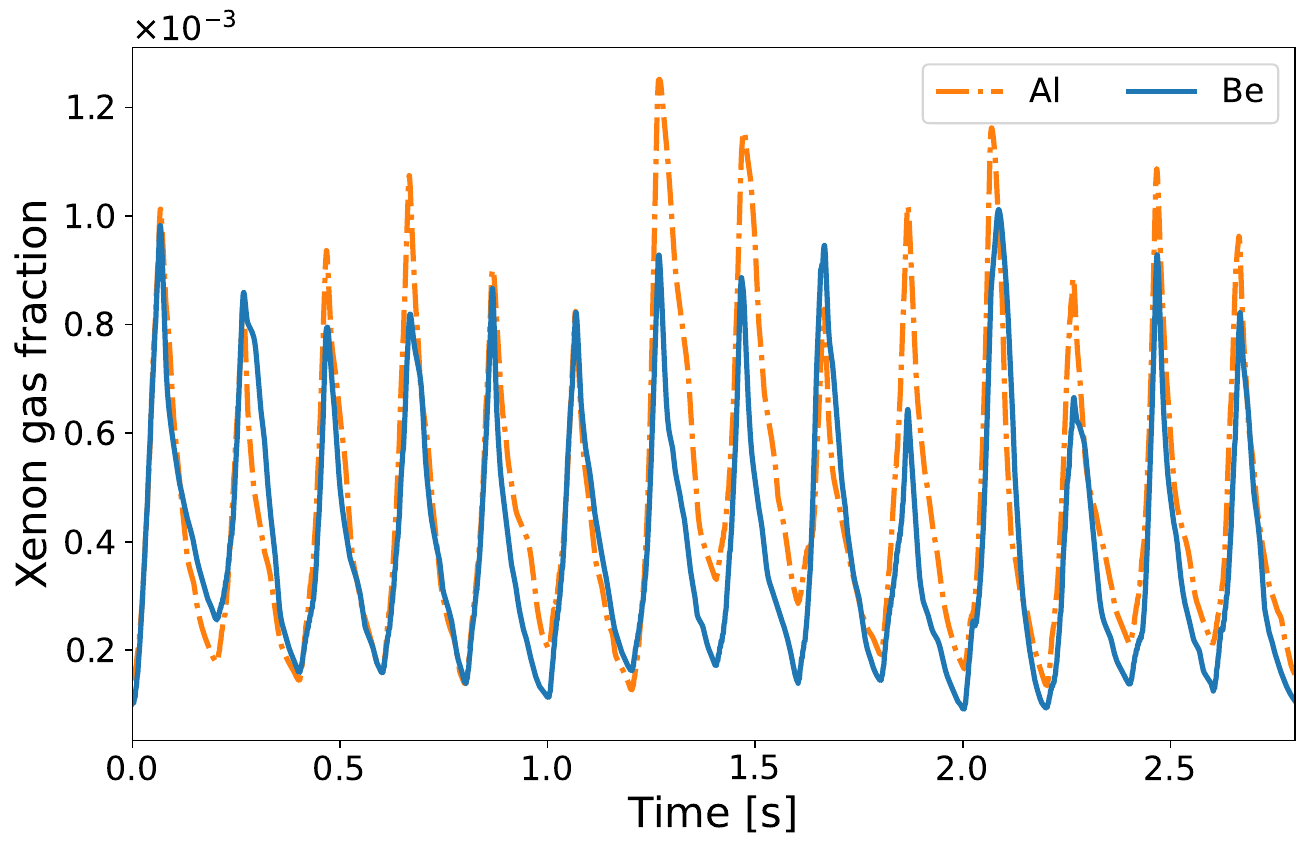}
    \caption{Fraction of xenon in the gaseous state for multiple bunch trains. Our simulations suggest that boiling mainly occurs at the exit window and is consistent with aluminum reaching higher temperatures than beryllium.}
    \label{fig:gasfrac}
\end{figure}

We find that xenon boiling is most prevalent at the exit window boundary, in which the LXe receives additional heating from the exit window itself. \cref{fig:gasfrac} illustrates the fraction of xenon gas in the LXe cell on the same time scale as in the other figures. In agreement with \cref{fig:exit_max_temp}, the hotter aluminum exit window results in greater LXe boiling when compared to beryllium. Based on the CFD simulations, the Be cell windows would be preferable, as they mitigate LXe boiling better than the Al cell windows. The next phase of cell design can be based on the CFD results to perform a mechanical stress and fatigue analysis accounting for the predicted temperature gradients in the cell's windows for both materials considered here, Al and Be.

\section{Conclusion}

A set of LXe positron target simulations were carried out in G4Beamline to obtain spatial distributions of the energy deposition. These results were used in ANSYS Fluent to create an optimized LXe flow structure to minimize heating in the windows.
We find that LXe vaporization is minimal and is concentrated on the beam axis and near the exit window. Further, the vaporized xenon can be sufficiently evacuated from the IP before the next bunch train arrives at the target.

A significant portion of the target exit window exceeds the PEDD at 0.5 mm thickness for both beryllium and aluminum. Here, we assume a 35 Jg$^{-1}$ maximum but acknowledge that further testing should be done verify this limit.
Regardless, a suitable window material must sustain the thermal shocks induced by the bunch train without losing structural integrity. Whether this is enabled via sufficiently low radiation lengths or through material properties that raise the PEDD has not yet been addressed.

In this work, we assume a pure LXe for the target. This is a reasonable assumption, as the EXO collaboration has managed to produce large scale Xe with less than 1~ppb impurities~\cite{Ackerman2022}. While many of the engineering challenges of using liquid xenon as a positron source remain unresolved, this work demonstrates that LXe offers an alternative, reusable, non-toxic approach towards realizing a positron target for an ILC-type collider.

\section*{Declaration of competing interest}

The authors declare that they have no known competing financial interests or personal relationships that could have appeared to
influence the work reported in this paper.

\section*{Data availability}

All data used in this study was obtained via G4Beamline~\cite{Roberts2007} and ANSYS Fluent~\cite{ansys_fluent_2025r1,ansys_fluent_2026r2}. The G4Beamline input scripts and data analysis files used to produce the results of this work are openly available online~\cite{source_code}, and the simulation data used in this work is publicly available on Zenodo~\cite{ZenodoData}.

\section*{Acknowledgments}

The authors are supported by the U.S. Department of Energy under contract DE–AC02–76SF00515.



\printcredits

\bibliographystyle{edited-elsarticle-num}

\bibliography{cas-refs}



\end{document}